\documentclass[twocolumn]{aastex6}
\usepackage{epstopdf}
\begin{document}


\title{CO$_2$ Infrared Phonon Modes in Interstellar Ice Mixtures}


\author{Ilsa R. Cooke\altaffilmark{1}}
\affil{Department of Chemistry \\
University of Virginia \\
McCormick Rd\\
Charlottesville, VA, 22904, USA
}

\and
\author{Edith C. Fayolle and Karin I. {\"O}berg}
\affil{Harvard-Smithsonian Center for Astrophysics\\
60 Garden Street\\
Cambridge, MA, 02138, USA}


\altaffiltext{1}{irc5zb@virginia.edu}

\begin{abstract}

CO$_2$ ice is an important reservoir of carbon and oxygen in star and planet forming regions. Together with water and CO, CO$_2$ sets the physical and chemical characteristics of interstellar icy grain mantles, including desorption and diffusion energies for other ice constituents. A detailed understanding of CO$_2$ ice spectroscopy is a prerequisite to characterize CO$_2$ interactions with other volatiles both in interstellar ices and in laboratory experiments of interstellar ice analogs.  We report laboratory spectra of the CO$_2$ longitudinal optical (LO) phonon mode in pure CO$_2$ ice and in CO$_2$ ice mixtures with H$_2$O, CO, O$_2$ components. We show that the LO phonon mode position is sensitive to the mixing ratio of various ice components of astronomical interest. In the era of JWST, this characteristic could be used to constrain interstellar ice compositions and morphologies. More immediately, LO phonon mode spectroscopy provides a sensitive probe of ice mixing in the laboratory and should thus enable diffusion measurements with higher precision than has been previously possible.
\end{abstract}

\keywords{Astrochemistry --- molecular processes ---
methods: laboratory: solid state}



\section{Introduction} \label{sec:intro}
In cold regions of the interstellar medium dust grains are coated with ices -- H$_2$O, CO, CO$_2$ and additional minor ice constituents -- through a combination of condensation and an active surface chemistry. These ices account for up to 60 and 80\% of the volatile oxygen and carbon budget in star forming regions respectively \citep{Oberg2011}. Of the three major ice constituents, this study focuses on CO$_2$. CO$_2$ is an important ice component during all stages of star formation. CO$_2$ ice has been observed in dense clouds \citep{Whittet1998,Bergin2005,Knez2005,Whittet2009,Noble2013}, protostellar envelopes \citep{D'Hendecourt1989,Boogert2004,Pontoppidan2008,Shimonishi2010,Aikawa2012} and in comets \citep{Ootsubo2012}, the remnants of the Solar Nebula. In this range of environments, CO$_2$ abundances with respect to water is surprisingly constant at 20--30\%.

Interstellar and cometary CO$_2$ identifications as well as column density determinations rely on infrared spectroscopy.
CO$_2$ ice has two IR active fundamental modes: the 4.27 $\micron$ stretch ($\nu_3$) and the 15.2 $\micron$  bending ($\nu_2$) modes, both of which have been identified in ices observationally. Furthermore, CO$_2$ ice spectra have been shown to depend sensitively on the local ice environment. The 15.2 $\mu$m band displays a characteristic Davydov splitting in the pure crystalline phase. Amorphous CO$_2$ ice and CO$_2$ in ice mixtures do not show this splitting, and the band is typically shifted with respect to the pure ice band position. The frequency and shape of the CO$_2$ $\nu_3$ stretch also depend on the bulk ice structure. Several studies have  pointed out that the band displays a low frequency shoulder at 2328 cm$^{-1}$ (4.30 $\mu$m) in pure amorphous CO$_2$ and in some CO$_2$ mixtures with hydrogen-bonding molecules \citep{Ehrenfreund1997,Gerakines2015,Escribano2013}. This environmental sensitivity has been used to identify several different CO$_2$ ice phases in observational spectra: in star forming regions, most CO$_2$ is typically mixed with water, but CO:CO$_2$  and pure CO$_2$ ice phases are also common \citep{Pontoppidan2008,White2009}. CO$_2$ ice spectroscopy has also been used in the laboratory to trace CO$_2$ trapping, segregation and diffusion processes \citep{Ehrenfreund1998,Ehrenfreund1999,Palumbo2000,Oberg2009}.   

In addition to the normal vibrational modes, CO$_2$ optical phonons, arising from long range collective vibrations in the solid, can be excited by infrared radiation. 
A phonon is a quantized vibrational motion in which the lattice atoms or molecules vibrate at a single frequency. Optical phonons occur when the molecules are moving out of phase within the lattice. These phonons can propagate through thin films of astronomical dimension but cannot travel through films of thickness much greater than the wavelength of incoming radiation. Thick films will instead exhibit Restrahalen bands in which the change in the refractive index results in a strong reflection.
Solid CO$_2$ exhibits both tranverse optical (TO) and longitudinal optical (LO) phonons in which the normal vibrations propagate through the ice lattice perpendicular and parallel to the direction of the IR field respectively.
When a thin film is positioned at an oblique angle to the incoming radiation, the electric field has components both parallel and perpendicular to the film normal and longitudinal optical phonons can be excited by the parallel component of the field vector.  
The angular dependent enhancement of the LO mode is known as the Berreman effect after experiments conducted by Berreman on the angular dependence of the LO mode in ionic crystals \citep{Berreman1963}. 

The theory of optical phonon mode splitting has been described previously for polycrystalline films \citep{Ovchinnikov1993}. The magnitude of the splitting is estimated by:

\begin{equation} \label{eq:1}
	\nu_{LO}^2 - \nu_{TO}^2 = \frac{\mathrm{4\pi}}{\varepsilon m V}
	\left(\frac{\partial\mu}{\partial q}\right)^2	
	\end{equation} 
where $\nu_{LO}$ and $\nu_{TO}$ are the frequencies of the LO and TO phonon modes, $\varepsilon$ is the dielectric constant of the ice, \textit{m} is the reduced mass associated with the coordinate \textit{q}, \textit{V} is the unit cell volume and $\partial\mu$/$\partial q$ is the transition dipole moment matrix element.  

In many cases the splitting can be accurately described by the Lyddane-Sachs-Teller (LST) relationship:

\begin{equation} \label{eq:2}
	\frac{\nu_{LO}}{\nu_{TO}} = \sqrt{\frac{\varepsilon_{0}}{\varepsilon_{\infty}}} = \frac{n_{0}}{n_{\infty}}
\end{equation}

where \textit{n}$_{0}$/$\varepsilon_{0}$ and \textit{n}$_{\infty}$ /$\varepsilon_{\infty}$ are the limiting low- and high-frequency refractive indices/ dielectric constants of the band. The high-frequency index is often approximated using the electronic dielectric constant, for example \citet{Hudgins1993} use reported values of the visible refractive index of the sample at the sodium D line ($\lambda$ = 589 nm). The values used by Hudgins \textit{et. al.} for CO$_2$, O$_2$, CO and H$_2$O are \textit{n}$_{\infty}$ = 1.22, 1.25, 1.30 and 1.32. 
The LST approximation was originally developed for cubic crystals but has been shown to apply to a range of disordered materials \citep{Whalley1977,Whalley1979,Sievers1990}.

LO phonon modes are often invisible in laboratory ice spectra due to the common experimental setup in which the infrared beam is normal to the film surface.  CO$_2$ LO-phonon modes have, however, been reported in Reflection-Absorption Infrared Spectroscopy (RAIRS) studies of ices, when the substrate is naturally positioned at a grazing angle to the IR beam.  \citet{Baratta1998} report RAIR spectra of CO$_2$ ices at 12 K and show that new modes appear using p-polarized light at 2377 and 676 cm$^{-1}$. 

\cite{Escribano2013} have also observed CO$_2$ LO phonons in both transmission IR and RAIRs experiments. They show that the CO$_2$ $\nu_{3}$ mode in p-polarized RAIR specta shifts towards the LO frequency of pure crystalline CO$_2$ ice (2381 cm$^{-1}$) with increasing ice thickness.

The CO$_2$  $\nu_{3}$ LO phonon mode has also been observed in ASW films exposed to CO$_2$ at 90 K \citep{Kumi2006} and for H$_2$O/CO$_2$ nanoparticles at 80 K \citep{Taraschewski2005}. CO$_2$ deposited on amorphous solid water (ASW) ice at 90 K exhibits an LO mode at 2379 cm$^{-1}$, close to that of pure CO$_2$ ice. Composite H$_2$O/CO$_2$ nanoparticles show shape effects which manifest in the IR at frequencies between $\nu_{LO}$ and $\nu_{TO}$. 

In this paper we present infrared transmission spectra of the CO$_2$ optical phonon modes in low temperature ices. We demonstrate that the LO phonon mode shape and peak frequency depends sensitively on ice mixing components, such as H$_2$O, CO and O$_2$. We analyze these dependencies and find a systematic frequency shift with ice mixing ratio for all mixing-partners, regardless of whether the ice mixture is deposited directly, or produced photolytically {\it in situ}. In section \ref{sec:discussion}, we discuss how these spectroscopic characteristic could be used to characterize CO$_2$ in astrophysical environments, as well as in laboratory ice experiments aimed at constraining ice diffusion.

\begin{figure}
\figurenum{1}
\plotone{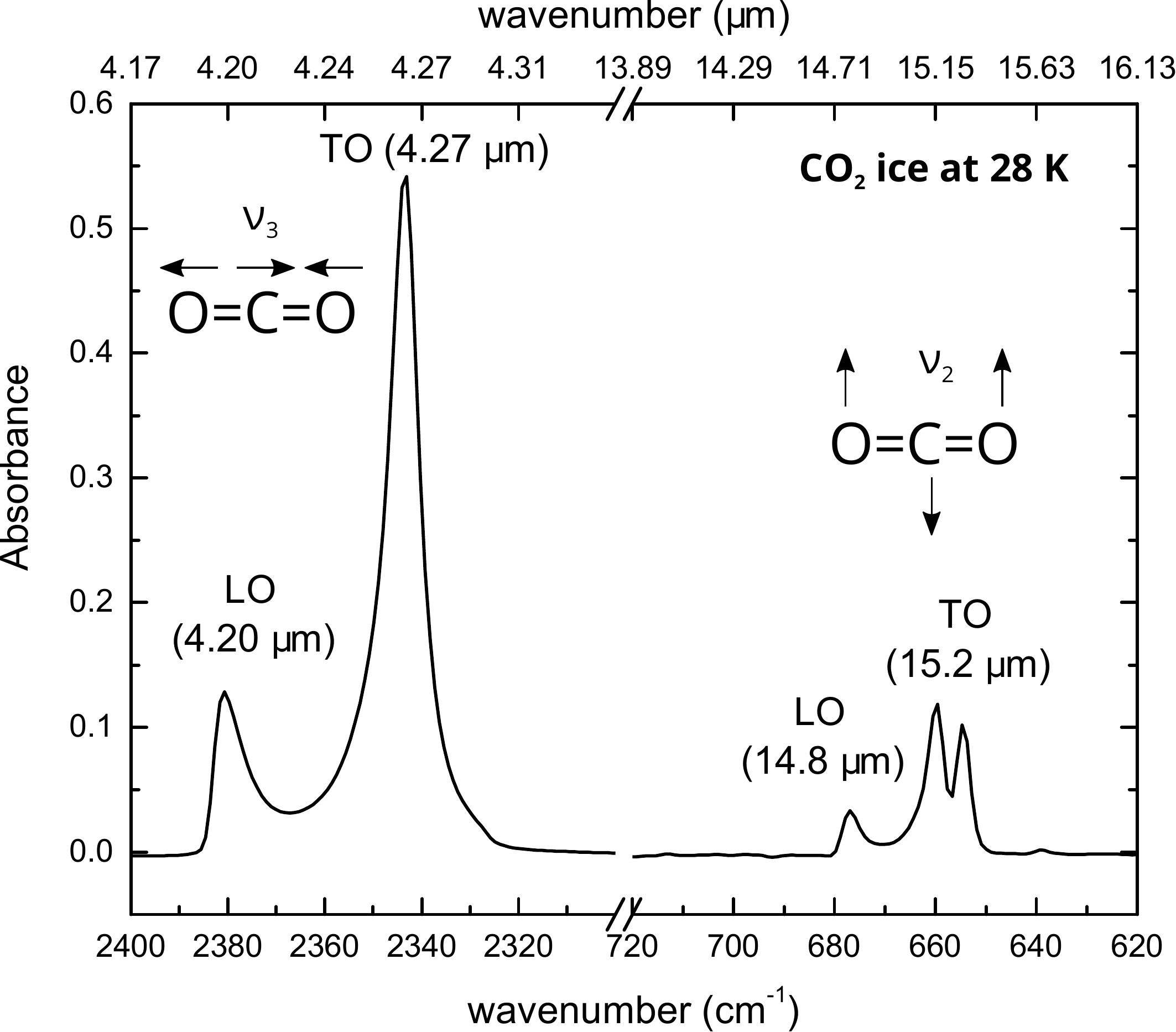}
\caption{Infrared spectrum of pure CO$_2$ ice deposited on KBr at 28 K at 40$^{\circ}$ grazing angle to the IR beam. The two strongest spectral features are shown, the CO$_2$ asymmetric stretch ($\nu_3$) at 4.27 $\mu$m and the CO$_2$ bend ($\nu_2$) centered around 15.2 $\mu$m. \label{fig:pureCO2}}
\end{figure}

\begin{figure}
	\figurenum{2}
	\plotone{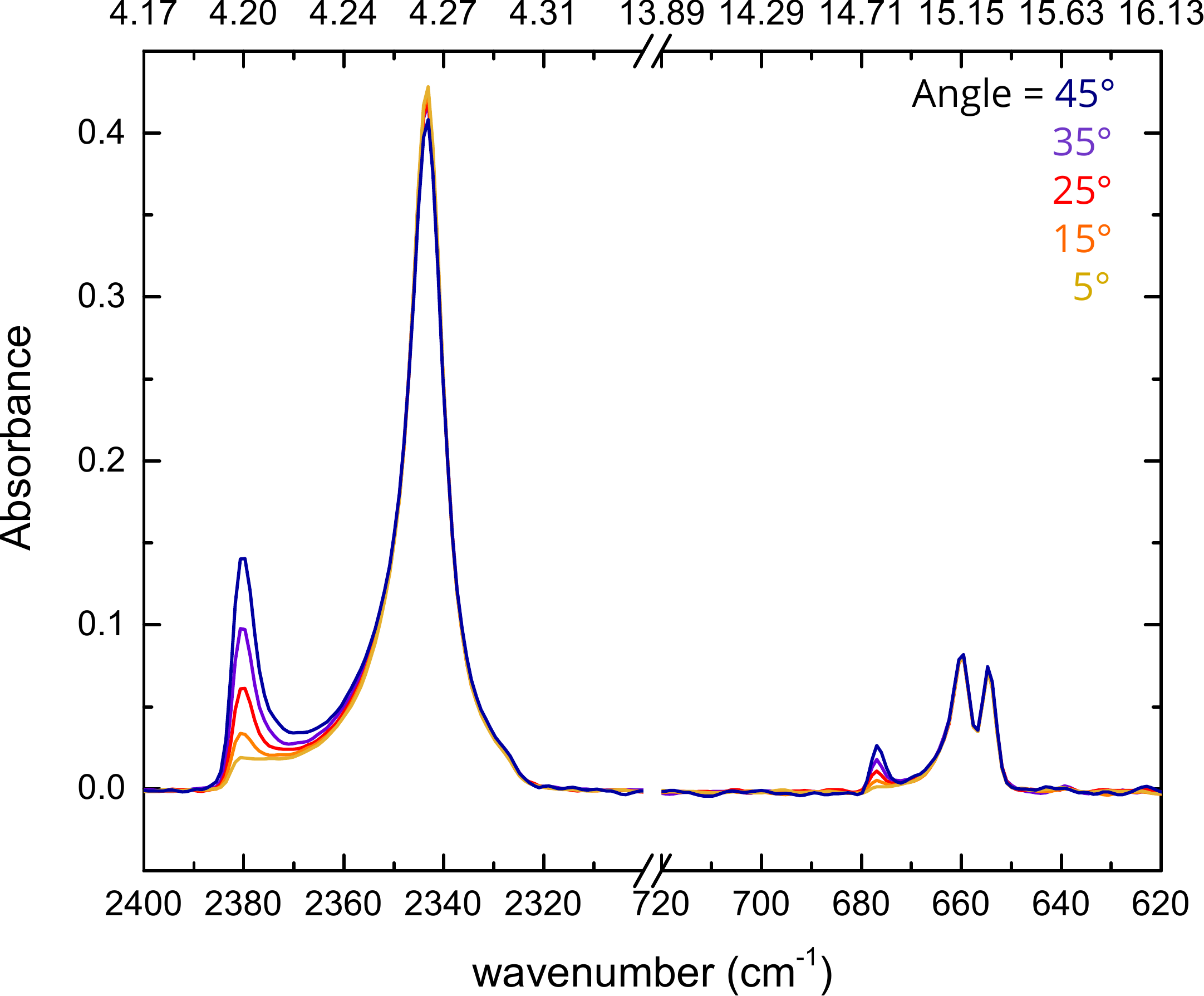}
	\caption{Angular dependence of the infrared spectrum of pure CO$_2$ ice deposited on KBr at 28 K. The IR spectra are shown for angles between 5-45$^{\circ}$ from the IR beam incidence. The LO phonon mode intensity decreases as the film is rotated toward the normal incident angle. \label{fig:anglesCO2}}
\end{figure}
\section{Methods} \label{sec:meth}

\begin{figure}
\figurenum{3}
\plotone{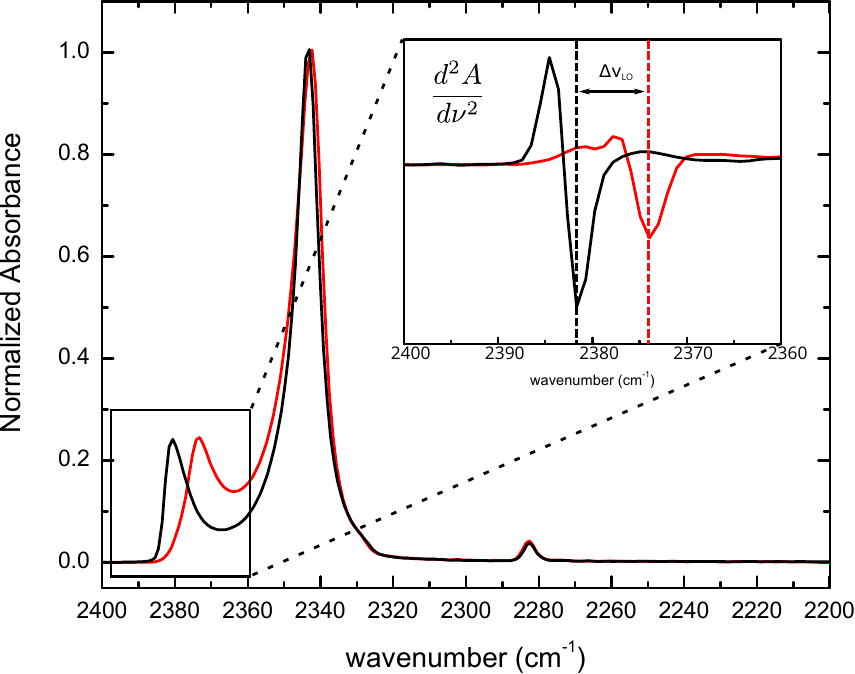}
\caption{Infrared spectra of pure CO$_2$ ice (black) and a CO$_2$:CO mixed ice (red). The inset shows the second derivative spectra that are used to determine the frequency of the LO phonon mode. The redshift of the LO phonon mode is calculated as the difference between the LO frequency in pure CO$_2$ and in the mixed ice.  \label{fig:deriv}}
\end{figure}

Laboratory ice studies were conducted in a high vacuum system (base pressure $\sim$2 x 10$^{-8}$ Torr, dominated by H$_2$) modified from \citet{Yuan2013}. A closed cycle helium cryostat (Air products Displex DE 202-0SP expander and APD Cryogenics HC-4 compressor) was added to cool the IR transparent substrate, a 0.3 cm$^{2}$ KBr pellet pressed into a tungsten grid that is suspended into the chamber by an OFHC copper coldfinger. The substrate is cooled to 28 K and the temperature is monitored using a k-type thermocouple welded to tantalum foil and attached to the W-grid close to the KBr pellet. Infrared spectra were collected using a Bruker Tensor 27 FTIR spectrometer and a liquid-N$_2$ cooled MCT detector. In the fiducial experiment the substrate was kept at $\sim$40$^{\circ}$ to the IR beam, but other angles were explored as well.   

H$_2$O was prepared in a He-purged glove-box and freeze-pump-thawed for several cycles. The purity of the CO$_2$, CO, O$_2$ and H$_2$O used to prepare the ice mixtures was checked prior to the experiments using an RGA 200 quadrupole mass spectrometer (QMS). 
Ice mixing ratios are calculated using infrared absorption spectra and literature band strengths \citep{Gerakines1995,Oberg2006} when possible, i.e. for H$_2$O, CO$_2$ and CO. Uncertainties on ice mixing fractions are calculated by analysis of repeat absorbance spectra of the ice mixtures and the pure ice, combined with a 15\% uncertainty in the bands strengths used to calculate the ice mixing fraction. Molecular oxygen is not observed in the infrared due to absence of a permanent dipole. Oxygen is mixed with CO$_2$ in the gasline and the measured O$_2$ gas fraction was used to approximate the ice mixture. The gas fraction was measured by monitoring the O$_2$ and CO$_2$ partial pressures with the QMS at m/z = 32 and 44 respectively. 

For photochemistry experiments, a microwave discharge hydrogen flow lamp (Opthos Instruments) was used. It has been described in detail previously \citep{Rajappan2010}. The microwave power was kept at 60 W for the duration of the experiments. A mixture of 90$\%$ Ar, 10$\%$ H$_2$ was used to produce UV light with a sharp Lyman-alpha feature \citep{Okabe1964}. 

\section{Results} \label{sec:results}

\begin{figure*}
	\figurenum{4}
	\gridline{\fig{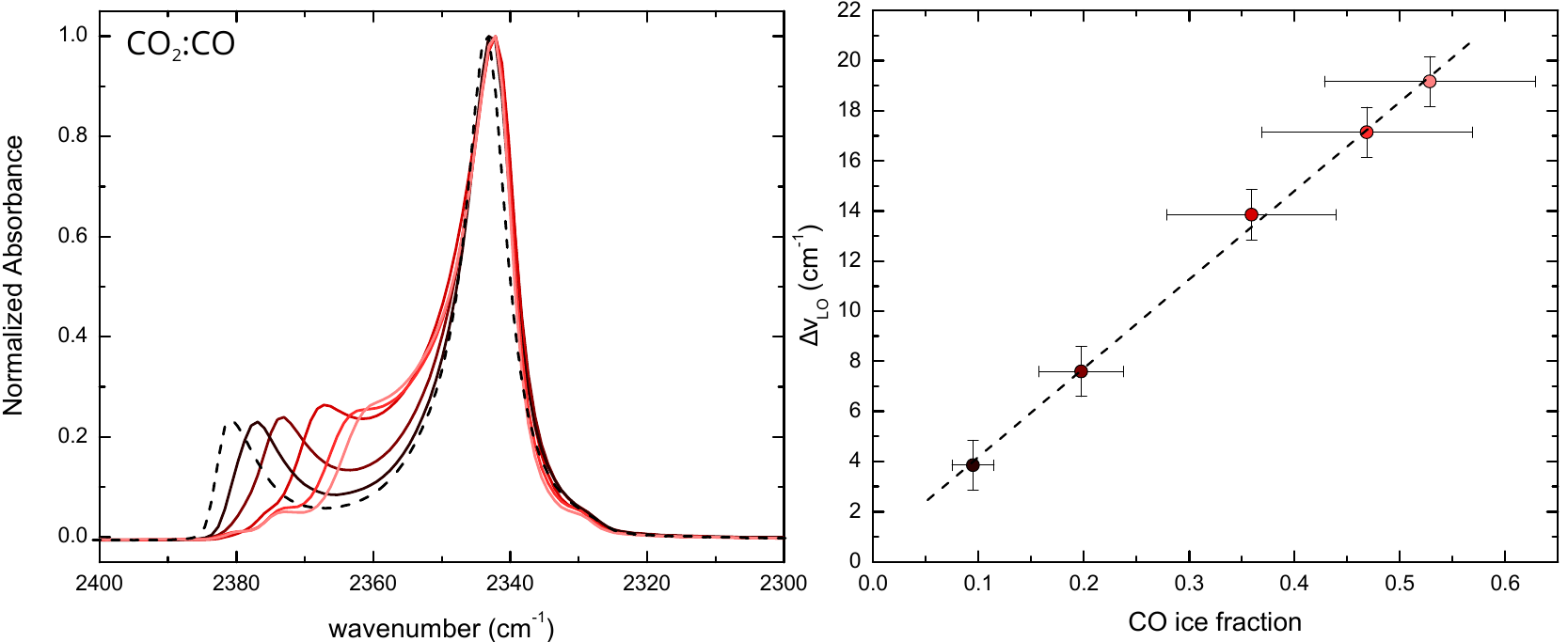}{0.6\textwidth}{}
	}
	\gridline{\fig{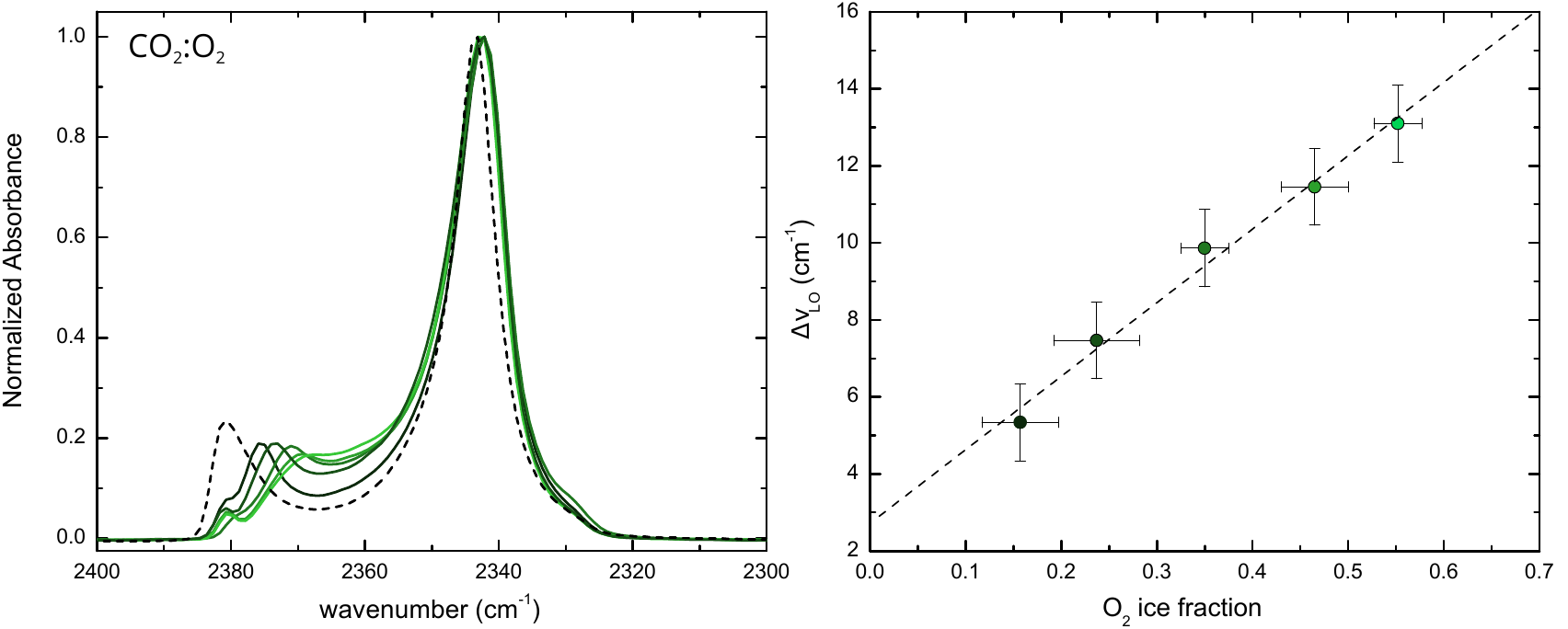}{0.6\textwidth}{}
	}
	\gridline{\fig{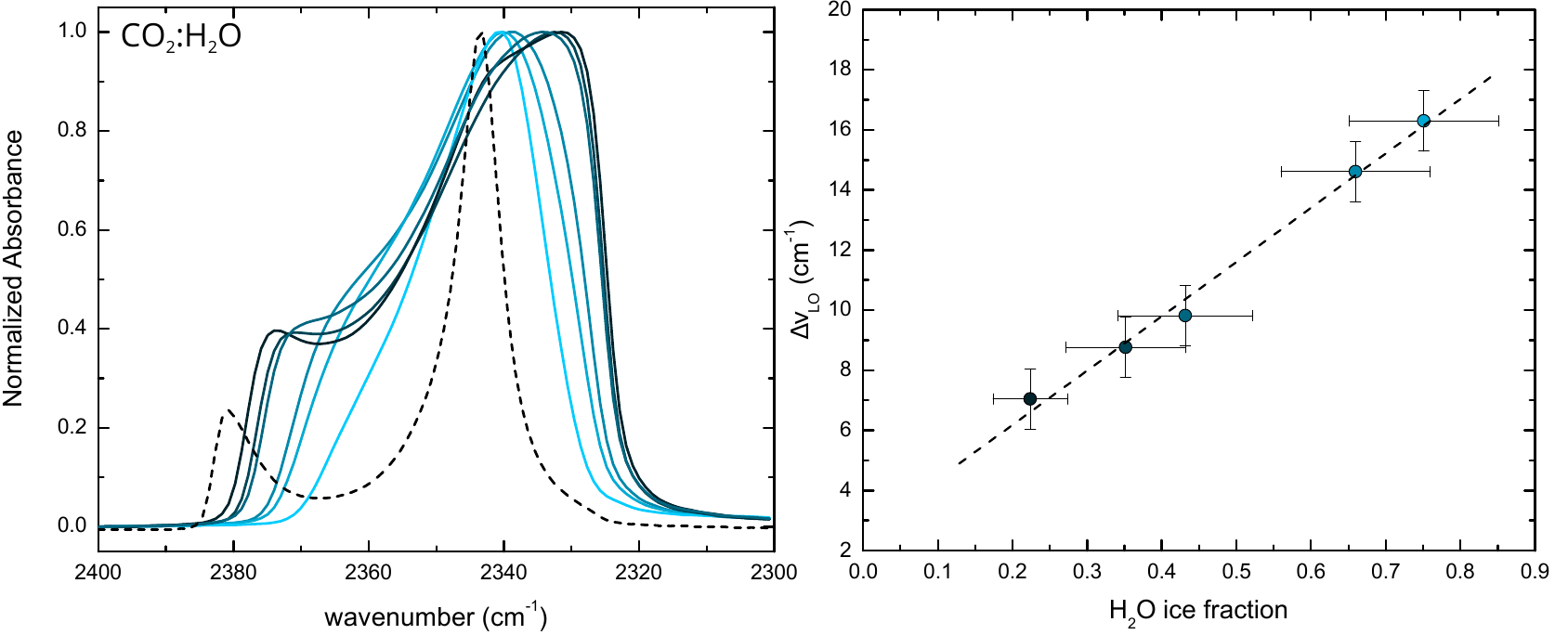}{0.6\textwidth}{}}
	
	\caption{\textit{Left}: Spectra of defect (top: CO, middle: O$_2$, bottom: H$_2$O) bearing CO$_2$ ices deposited from gas mixtures at 28 K. The dashed line shows the spectrum of pure CO$_2$ ice. The CO$_2$ LO phonon mode redshifts with increasing defect fraction in the ice. \textit{Right}: Shift of the LO phonon mode with respect to pure CO$_2$ ice for various concentrations of the defects in CO$_2$ ices. Errors on the ice fraction are $\pm$1$\sigma$ on the mean calculated from repeated spectral measurements of the ice film in addition to a 15\% error on the band strengths used to calculate the ice column densities. Oxygen fractions and errors are based on QMS measurements of the gas deposition mixture that have been adjusted for surface sticking. Errors on the LO shift are calculated from the spectral resolution.
		\label{fig:spectra}}
\end{figure*}

Figure \ref{fig:pureCO2} shows the infrared spectrum of pure CO$_2$ ice at 28 K, taken with the substrate at 40$^{\circ}$ to the incident IR beam. The LO phonon mode appears at 2381 cm$^{-1}$ (4.2 $\mu$m) and is shifted from the $\nu_3$ (TO) mode by 39 cm$^{-1}$. 
The CO$_2$ $\nu_2$ mode also exhibits an LO phonon mode at 677 cm$^{-1}$ (14.8 $\mu$m), split 17 cm$^{-1}$ and 22 cm$^{-1}$ from the CO$_2$ doublet peaks at 660 and 655 cm$^{-1}$ respectively. The LO phonon mode disappears as the ice film is rotated such that its plane of incidence is parallel to the IR beam (figure \ref{fig:anglesCO2}). {\color{black}The CO$_2$ ices deposited at 28 K have crystalline spectral characteristics including splitting of the $\nu_2$ mode and both the frequency and bandwidth of $\nu_3$ mode, as well as an amorphous feature around 2328 cm$^{-1}$ that dominates at low coverage.} The CO$_2$ ices in each of the experiments are 70-90 nm thick, calculated by integrating the $\nu_1$ + $\nu_3$ mode using the band strengths of \citet{Gerakines1995} and a density of 1.78 g/cm$^3$ as reported for crystalline CO$_2$. 

Figure \ref{fig:deriv} shows the method for determining the $\Delta\nu_{LO}$ by analysis of spectra of the ice mixtures and pure CO$_2$ ices.  The $\nu_{LO}$ frequency is found by identifying the minimum in the second derivative spectrum. Second derivative spectra were taken in both the Opus software and Thermo Scientific Grams/AI Spectroscopy software. A smoothing function was applied in Opus software before taking the second derivative. The frequencies of the $\nu_{LO}$ modes from the smoothed spectra in Opus agreed within 1\% with those determined using Grams without smoothing. The uncertainty on $\Delta\nu_{LO}$ is conservatively taken as the spectral resolution ($\pm$ 1 cm$^{-1}$). 

\begin{deluxetable}{ccccc}
	\tablecaption{Measured CO$_2$ LO frequencies for CO$_2$ ice mixtures \label{tab:mixing}}
	\tablecolumns{4}
	\tablenum{1}
	\tablewidth{0pt}
	\tablehead{
		\colhead{Ice} &
		\colhead{Mixing Fraction} &
		\colhead{$\nu_{LO}$} &
		\colhead{$\Delta\nu_{LO}$} \\
		\colhead{} &  \colhead{} & \colhead{(cm$^{-1}$)} &
		\colhead{(cm$^{-1}$)}
	}
	\startdata
	CO$_2$:CO & 0.10$\pm$0.02 & 2377.0 & 3.9 \\ 
	& 0.20$\pm$0.04 & 2373.3 & 7.6\\
	& 0.36$\pm$0.08 &  2367.0 & 13.8\\
	& 0.47$\pm$0.10 & 2363.7 & 17.1\\
	& 0.53$\pm$0.10 & 2361.7 & 19.2\\ \hline
	CO$_2$:O$_2$  & 0.16$\pm$0.04 & 2375.5 & 5.3\\
	& 0.24$\pm$0.05 & 2373.4 & 7.5\\
	& 0.35$\pm$0.03 & 2371.0 & 9.9\\
	& 0.47$\pm$0.04 & 2369.4 & 11.5\\
	& 0.55$\pm$0.03 & 2367.8 & 13.1\\ \hline
	CO$_2$:H$_2$O & 0.22$\pm$0.05 & 2373.8 & 7.0\\
	& 0.35$\pm$0.08 & 2372.1 & 8.8\\
	& 0.43$\pm$0.09 & 2371.0 & 9.8\\
	& 0.66$\pm$0.10 & 2366.3 & 14.6\\
	& 0.75$\pm$0.10 & 2364.6 & 16.3\\
	\enddata
	
\end{deluxetable}

The CO$_2$ experimental spectra and $\nu_{3}$ LO mode redshifts versus defect ice fraction for CO, O$_2$ and H$_2$O ice mixtures are displayed in Figure \ref{fig:spectra} and Table \ref{tab:mixing}.
Figure \ref{fig:spectra} shows the effect of introducing CO, O$_2$ or H$_2$O into the CO$_2$ ice lattice by mixing with CO$_2$ prior to deposition. Increasing the fraction of the defect species in the CO$_2$ ice results in an increasing redshift of the LO phonon mode from the pure ice position. The right panels of figure \ref{fig:spectra} present the $\nu_{LO}$ redshift versus the fraction of each species in the CO$_2$ ice mixture.  These plots display a positive linear correlation between the defect concentration and the LO redshift for each of the ice mixtures considered. The slope of the $\nu_{LO}$ redshift versus the defect fraction are 36 $\pm$ 7 cm$^{-1}$ N$_{CO}$/N$_{total}$, 19 $\pm$ 4 cm$^{-1}$ N$_{O_2}$/N$_{total}$ and 18 $\pm$ 4 cm$^{-1}$ cm$^{-1}$ N$_{H_2O}$/N$_{total}$ for CO, O$_2$ and H$_2$O ice mixtures respectively. The effect of the ice defect on the position of the LO mode may be quantified by comparing these slopes. For example, we would expect the LO mode to shift by 1 cm$^{-1}$ if the CO$_2$ ice contains $\sim$3$\%$ CO, $\sim$5$\%$ O$_2$ or $\sim$6$\%$ H$_2$O.  
 
In addition to shifting the frequency of the LO mode, the three mixing partners also affect the shape of the LO mode spectral band, and these spectral changes are species specific.
In CO mixtures with CO$_2$, a second small peak appears in the spectra of mixtures with more than 30$\%$ CO. This peak is also redshifted from the pure LO phonon mode position but does not appear to shift with CO concentration beyond 30\%, possibly representing a stable mixing environment between CO and CO$_2$ such as the formation of a complex. 

A similar second peak is present in all of the mixed O$_2$:CO$_2$ ice spectra. 
The spectra of H$_2$O mixtures with CO$_2$ differ from those of the CO and O$_2$ mixtures in several ways.  
In particular, the spectra of CO$_2$ in polar H$_2$O environments show broadening of the $\nu_3$ mode as well as non-linear dependence of the $\nu_3$ frequency on the fraction of H$_2$O in the ice. 

\begin{figure*}[t!]
	\figurenum{5}
	\includegraphics[scale =0.7]{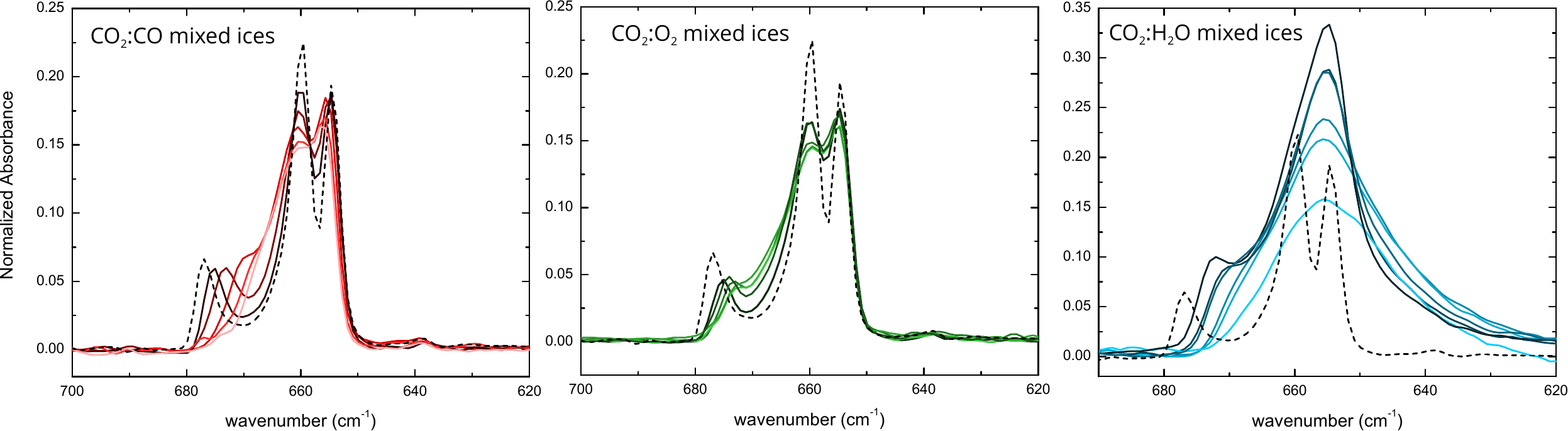}
	\caption{CO$_2$ bending modes for 
		CO:CO$_2$ mixtures, O$_2$:CO$_2$ mixtures and H$_2$O:CO$_2 $ mixtures \label{fig:v2}. The colors correspond the mixing ratios shown in Figure \ref{fig:spectra}}
\end{figure*}

The CO$_2$ bending mode for the ice mixtures are shown in Figure \ref{fig:v2}. The LO phonon mode is split from the $\nu_{2b}$ mode at 660 cm$^{-1}$ by 17 cm$^{-1}$, while the $\nu_{2a}$ peak at 655 cm$^{-1}$ does not display appreciable LO/TO splitting. The $\nu_2$ LO mode behaves similarly to the $\nu_3$ mode, with linear shifts that are dependent on the fractional abundance of the defect in the ice. The redshift is likewise the largest for CO ice mixtures with respect to the ice fraction.  Both CO and O$_2$ ice mixtures display decreased Davydov splitting of the $\nu_2$ band with respect to the pure CO$_2$ ices. The $\nu_2$ bands in H$_2$O mixtures are broad and do not display Davydov splitting, due to the effect of the hydrogen bonding  network of the H$_2$O lattice on the CO$_2$ crystal structure. The decrease of Davydov splitting in the ice mixtures supports the fact that the CO$_2$ crystal structure is disrupted upon addition of the defect molecules as pairwise binding interactions between CO$_2$ and the introduced molecule can result in only one equivalent CO$_2$ per unit cell. This is consistent with the fact that the Davydov splitting continues to decrease in the ice mixtures with increasing concentration of CO or O$_2$ as more interactions become available in each unit cell.

\begin{figure*}
\figurenum{6}
\plotone{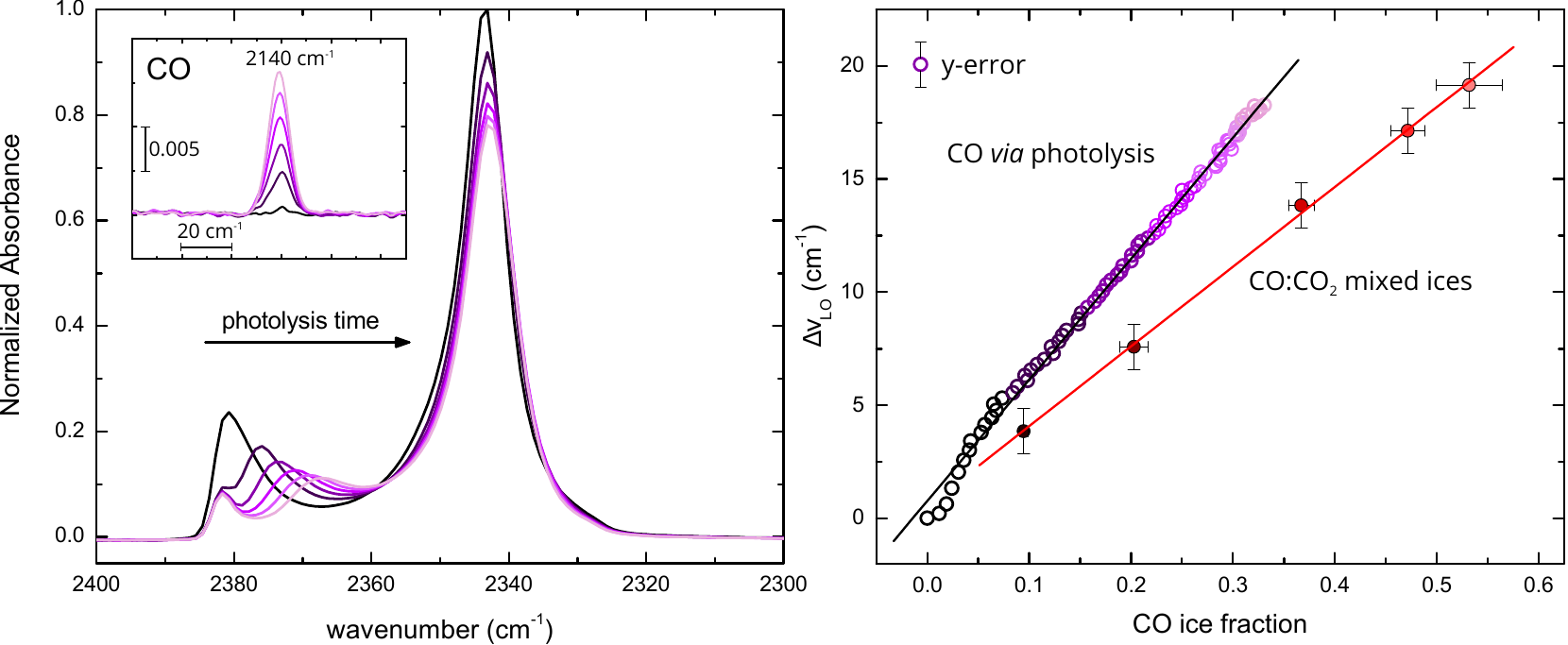}
\caption{Photolysis of a pure CO$_2$ ice using 10.2 eV photons. The CO$_2$ ice spectra show redshifts in the LO phonon mode that correlate to the linear production of CO photoproducts (inset) and depletion of CO$_2$ ice. The linear shift of the LO mode for CO$_2$:CO mixed ices is shown in red for comparison. \label{fig:photo}}
\end{figure*}

\begin{figure*}
\figurenum{7}
\plotone{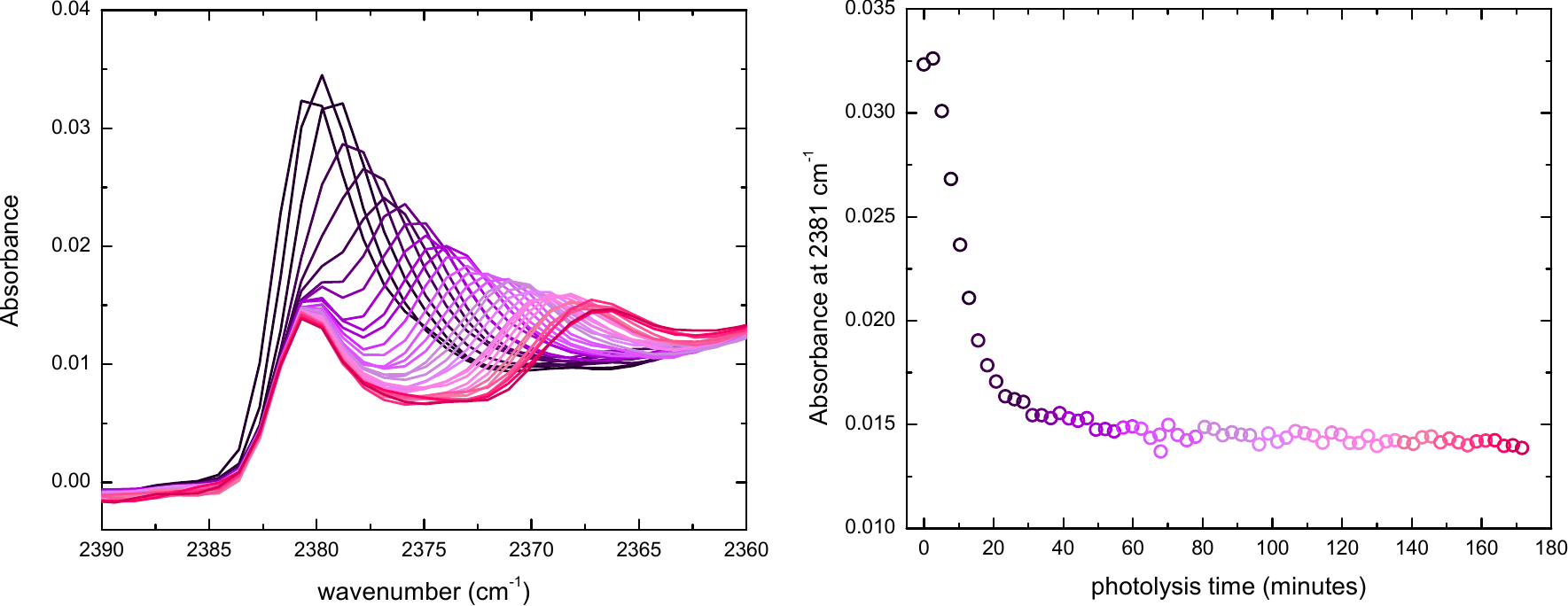}
\caption{Photolysis of a pure CO$_2$ ice $\sim$10 nm thick using 10.2 eV photons. The thin CO$_2$ ice spectra show redshifts in the LO phonon mode similar to that seen for the thicker ice. The RHS shows the absorbance at 2381 cm$^{-1}$ over the irradiation period. The residual phonon peak at 2381 cm$^{-1}$ plateaus after $\sim$20 minutes of irradiation.  \label{fig:photothin}}
\end{figure*}

We also conducted experiments to measure the introduction of defects \textit{in situ} by UV irradiation of pure CO$_2$ ices, which is known to produce CO. Figure 
\ref{fig:photo} shows spectra of pure CO$_2$ ice irradiated with UV 
photons from a hydrogen discharge lamp for a period of 140 minutes. The 
CO$_2$ LO phonon mode clearly redshifts with exposure time. The redshift is linear with CO photoproduct growth, similar to when a defect is introduced through pre-mixing. There are two differences in the LO phonon mode shifts in these experiments compared to the pre-mixed experiments, however. {\color{black}First, the LO phonon mode depends more steeply on CO concentration in the ice with a linear fit of 54 $\pm$ 6 cm$^{-1}$ N$_{CO}$/N$_{total}$ compared to the pre-mixed case which gives a fit of 36 $\pm$ 7 cm$^{-1}$ N$_{CO}$/N$_{total}$.}
Second, the photolyzed phonon mode is split into the redshifted (CO:CO$_2$ mixture) component and a second component that remains at the pure CO$_2$ LO phonon frequency. This suggests that some pure CO$_2$ is present throughout the photolysis experiment. To test whether this may be due to limited Ly-$\alpha$ penetration depth through the ice, we also photolysed a $\sim$10 nm think CO$_2$ ice while monitoring the LO phonon mode; the penetration depth of 10.2 eV photons through CO$_2$ ice has been previously measured in our laboratory as 50 nm \citep{Yuan2013}. Figure \ref{fig:photothin} shows that photolysis of the thin ice also results in a split LO phonon mode, with one peak that can be ascribed to pure CO$_2$ and another to a CO:CO$_2$ ice mixture, demonstrating that the second peak cannot be due to a limited photon penetration depth. 
Furthermore, the absorbance at the pure CO$_2$ phonon mode frequency (2381 cm$^{-1}$) plateaus after $\sim$20 minutes of irradiation, while the production of CO and destruction of CO$_2$ remains linear through the irradiation period. The plateau at the pure CO$_2$ LO frequency thus indicates that an environment exists in the ice in which the CO$_2$ is not mixed with the photoproducts and this hypothesis is further discussed in section \ref{sec:effects}.
 


\section{Discussion} \label{sec:discussion}

\subsection{Effects of CO, O$_2$ and H$_2$O on the CO$_2$ LO phonon mode} \label{sec:effects}

The LO-TO splitting depends on long range dipole interactions between CO$_2$ molecules and therefore is perturbed by the addition of other species into the ice lattice. The magnitude of the LO-TO splitting depends on the ice physical structure and bonding, the optical properties of the ice and the transition dipole moment of the CO$_2$ normal modes. Thus, the LO redshift upon addition of CO, H$_2$O and O$_2$ into the CO$_2$ ice lattice is a combination of how each of these species changes structure, binding energies, dielectric constant and transition moments of the CO$_2$ molecules within the lattice. 

According to equation \ref{eq:1}, the transition dipole moment governs the LO-TO splitting in polycrystalline films. The CO$_2$ asymmetric stretch has a large transition dipole matrix element, consistent with the large LO-TO splitting in pure CO$_2$ ice. One might expect that differences in the redshifts of the CO$_2$ LO frequency correspond to differences in the effects of the transition dipole elements of CO, H$_2$O and O$_2$ on that of the CO$_2$ LO mode. CO, while exhibiting only a small dipole moment, has a very large transition dipole moment that could perturb the effective CO$_2$ transition element. While H$_2$O has a large transition dipole, O$_2$ has $\mu$=0 for all internuclear separations and therefore its transition dipole moment is zero everywhere. If the changes in the CO$_2$ transition dipole moment were the only explanation for the LO redshift we would expect that O$_2$ would not shift the LO frequency and that CO and H$_2$O would have similar LO redshifts, which is not the case in our measurements. We therefore consider other effects on the LO-TO splitting. 

The addition of CO, O$_2$ and H$_2$O affects the CO$_2$ ice physical structure. Changes in the crystal structure can shift the LO mode to lower frequencies as shown for different CO$_2$ ice nanostructures \citep{Taraschewski2005}. Our data suggest that CO has the greatest effect on the longitudinal phonon frequency in CO$_2$ ice. Despite the substantially lower binding energy to CO$_2$ compared to H$_2$O, CO may produce the observed large shift because it is able fit well into the CO$_2$ ice lattice so that each CO$_2$ molecules in the lattice has dipole interactions with more CO molecules than in the case of H$_2$O; however, detailed structural simulations are required to constrain the exact mechanism. 

Splitting of the LO phonon into more than one band indicates that the mixing is not homogeneous and distinct CO$_2$ bulk environments are present in the ice. This may be caused by CO$_2$ segregation away from the defect ice component. Figure \ref{fig:spectra} shows splitting of the LO mode in the CO$_2$:O$_2$ ice mixtures and in the CO$_2$:CO  mixtures with high CO abundance. In the H$_2$O mixtures, the LO mode at the pure CO$_2$ frequency is absent, this suggests that the CO$_2$ is mixed in the H$_2$O ice lattice such that pockets of pure CO$_2$ do not occur. This is expected as CO$_2$ diffusion through H$_2$O is slow at 28 K \citep{Oberg2009} while diffusion and segregation of CO and O$_2$ may be possible, though further experiments are needed to support this hypothesis. 

We see the clearest splitting of the LO mode in the photolysis of CO$_2$ ice, where one component stays at the pure CO$_2$ LO position. As demonstrated by the thin ice photolysis experiment, this is not due to shielding of the UV radiation for the deepest ice layers. Instead, pockets of pure CO$_2$ must either be maintained or produced during the photolysis. Only $\sim$20$\%$ of the CO$_2$ ice is lost during the photolysis period. It is possible that some of the leftover CO$_2$ is not mixed with the photoproducts and maintained as pure CO$_2$ ice. Pure CO$_2$ could also be produced during the photolysis by several mechanisms. Excited CO$_2$ that is not photolysed carries excess energy that may allow it to segregate away from the photoproducts to form pure CO$_2$ pockets. CO$_2$ is also reformed during the photolysis by CO + O recombination and CO$_3$ dissociation. Another possible explanation is the desorption of photoproducts that are formed in the surface layers, leaving behind pure CO$_2$. This hypothesis could be investigated further by studying the molecules desorbing from the ice during the photolysis, by using isotopically labeled layers or by capping the CO$_2$ ice with an inert gas matrix.

\subsection{Potential Laboratory Applications of Phonon Mode Spectroscopy}

The sensitivity of the CO$_2$ LO phonon mode to ice mixing may be utilized in laboratory studies to provide information on bulk diffusion, physical mixing properties and ice phase properties such as crystallization. 
We have shown that the LO mode is particularly sensitive to the addition of ice defect molecules added either by deposition of gas mixtures or \textit{in situ} by photolysis of the ice. Mixing may also occur through diffusion of molecules in the ice. Diffusion is proposed to underpin much of the chemistry in interstellar ices, but diffusion barriers and mechanisms are highly uncertain \citep{Garrod2008,Lauck2015}. Bulk diffusion may also regulate the morphology of interstellar ices.  
Diffusion of CO out of CO$_2$ ice may cause CO distillation from CO-rich ices protostellar regions, thus explaining the presence of pure CO$_2$ in some such lines of sight \citep{Pontoppidan2008,Escribano2013}.  

We have conducted preliminary experiments to check whether the CO$_2$ LO mode could trace diffusion and desorption processes. The CO$_2$ ice mixtures were slowly heated passed the desorption temperature of CO or O$_2$ and the changes in LO mode were monitored. As the defect molecule desorbs, the LO phonon mode moves back toward the pure CO$_2$ frequency. The process is not completely reversible as some defect molecules remain trapped and desorb with the CO$_2$ ice around 75 K. This suggests that phonon mode spectroscopy could be used to characterize diffusion in CO$_2$ ices. Furthermore, phonon modes are not unique to CO$_2$ ice, and optical phonons in other ices may be used to measure diffusion coefficients for a range of astronomically relevant species. 

Phonon mode monitoring may also be used to characterize ice crystallization. The Lyddane-Sachs-Teller relationship (equation \ref{eq:2}) shows that the LO phonon mode should shift abruptly upon a phase change in the ice due to a rapid change in the static dielectric constant. Optical phonons in ices may therefore provide a useful measure of crystallization kinetics. The observed LO-TO splitting for ice deposited at different temperatures may be used to accurately deduce phase change temperatures and the corresponding index of refraction for ices around the phase change boundary.

\subsection{Astrophysical Implications} \label{sec:impl}

Longitudinal optical phonon modes may contribute to interstellar ice spectra for several important ice constituents if light from the background source is sufficiently polarized. In laboratory reflection absorption spectra p-polarized light, parallel to the surface normal, can interact with longitudinal phonons in the ice. In astronomical environments longitudinal phonons can be excited by p-polarized light or by the component of the unpolarized light wave that is parallel to the surface normal. Because the interstellar ices are not confined to slabs, as in the case of laboratory experiments, the fraction of p-polarization and s-polarization is generally equal for randomly orientated dust particles. Unpolarized light will predominantly produce the TO mode, though it can also excite LO phonons when the incoming IR wave is at oblique angles to the ice surface as seen in our laboratory spectra (Figure \ref{fig:anglesCO2}). The best sources for observing longitudinal phonons in ices are therefore those with the highest linear polarization fraction. Linear polarization arises from absorption and scattering from irregular dust grains which will absorb light from the background star preferentially along one axis. Dust grains can align with a local magnetic field and produce elliptically polarized light with linear and circular components \citep{Lazarian2007}. The degree of linear polarization in star forming regions is often very large ranging from 20-70\% and can extend over regions as large as 1 pc \citep{Kwon2013,Kwon2014}. A good source to characterize interstellar CO$_2$ LO phonon modes would be, for example IRAS 05329-0505, for which \citet{Kwon2014} measure the linear polarization fraction as 70\% in the K$_s$ band.

Optical phonon modes have been previously identified in laboratory spectra of H$_2$O and CO \citep{Whalley1977,Whalley1979,Baratta1998}. Water ice phonons are considered carriers of far infrared bands at 62 $\mu$m (longitudinal acoustic) and 44 $\mu$m (transverse optical) bands \citep{1990ApJ...355L..27O,Dartois1998}, which can be observed with far-IR telescopes such as SOFIA. These bands are sensitive to the ice structure with the 62 $\mu$m appearing at the crystallization phase transition \citep{Moore1992}. Spectra of CO bearing ices in 39 YSOs observed using the VLT have been fit using the optical phonon modes of $\alpha$-CO \citep{Pontoppidan2003}. Here, the LO mode was able to accurately reproduce the blue wing of the CO band for all of the observed sources. LO phonon modes thus needs to be taken into account when identifying the carriers of observed ice spectral features.

In the case of CO$_2$, identifying and characterizing the LO phonon mode in observed spectra could also provide important information about the ice composition, morphology and thermal history. Because the LO phonon mode is sensitive to defects in the CO$_2$ ice lattice, its frequency can be used to provide information about abundance of other species mixed with solid CO$_2$. In pristine CO$_2$ ice the LO mode will appear at 4.2 $\mu$m, if light from the background source is sufficiently polarized. For CO$_2$ ice in apolar and polar phases the LO phonon mode would be present at a redshifted frequency that is dependent on the fractional abundance of other ice components.  
A lack of LO-TO splitting in all ice species provides constraints on the source (polarization and dust properties), while a lack of LO-TO splitting in CO$_2$ when present in CO would constrain the CO$_2$ ice to be dilute, in which case the LO mode is convoluted with the TO mode.

To fit observational spectra, the CO$_2$ LO phonon spectra reported here need to be adjusted to account for grain size and shape effects. The standard method for treating grain shape and size variations has been to use a continuous distribution of ellipsoids (CDE) (e.g. \citet{Bohren83}). \citet{Ehrenfreund1997} have shown that the size and shape of grains affect the CO$_2$ absorption profile causing typical shifts and broadenings of 5 cm$^{-1}$ that can dominate over variations due to ice matrix effects. An in depth analysis of grain geometric effects on the CO$_2$ LO phonon mode will be needed to before it can be used to constrain ice mixing in observational spectra.
Analysis of LO-TO splitting should be especially fruitful on future ice spectra acquired with the JWST, since the analysis requires both high sensitivity and spectral resolution.

\section{Conclusions}

We have shown that longitudinal optical phonons in CO$_2$ ices are particularly sensitive to the presence of other ice species in the lattice. The frequency of the CO$_2$ LO mode redshifts linearly as the concentration of the ice mixing molecule is increased. Similarly, photolysis of CO$_2$ ices results in a linear redshift of the LO phonon mode with formation of the major photoproduct, CO. These spectral signatures will be useful for constraining ice mixing both in laboratory experiments and in observations of astronomical ices. \\

I. R. C. acknowledges full fellowship support from the Department of Chemistry, University of Virginia. I. R. C. would like to acknowledge instrument design and machining by Matthew Reish and helpful conversations with Rob Garrod, Eric Herbst, Patrick King and Jennifer Bergner.
I. R. C would like to acknowledge Jeffrey Shabanowitz for providing equipment used in producing these data. 
This work was conducted in the laboratory of the late John T. Yates Jr. to whom we would like to thank for his experimental techniques, motivation and mentorship leading to this study. 

%



\end{document}